\documentclass[aps,prl,twocolumn,superscriptaddress,showpacs]{revtex4}
\usepackage[T1]{fontenc}
\usepackage[latin1]{inputenc}
\usepackage{amsmath}
\usepackage{graphicx}
\usepackage{amssymb}

\makeatletter
\usepackage{subfigure}
\usepackage{amsfonts}

\makeatother
\begin{document}

\title{Probabilistic Phase Space Trajectory Description for Anomalous
  Polymer Dynamics}

\author{Debabrata Panja}
\affiliation{Institute for Theoretical Physics, Universiteit van
Amsterdam, Science Park 904, Postbus 94485, 1090 GL Amsterdam, The
Netherlands} 

\date{\today}

\begin{abstract} It has been recently shown that the phase space
  trajectories for the anomalous dynamics of a tagged monomer of a
  polymer --- for single polymeric systems such as phantom Rouse,
  self-avoiding Rouse, Zimm, reptation, and translocation through a
  narrow pore in a membrane; as well as for many-polymeric system such
  as polymer melts in the entangled regime --- is robustly described
  by the Generalized Langevin Equation (GLE). Here I show that the
  probability distribution of phase space trajectories for all these
  classical anomalous dynamics for single polymers is that of a
  fractional Brownian motion (fBm), while the dynamics for polymer
  melts between the entangled regime and the eventual diffusive regime
  exhibits small, but systematic deviations from that of a fBm.
\end{abstract}


\maketitle

In its terminal relaxation time $\tau$, $\tau\sim N^\kappa$, a polymer
of length $N$ displaces itself in space by its own size, which itself
scales as $\sim N^\xi$ \cite{degennes,de}. The values of $\kappa$ and
$\xi$ vary from system to system. E.g., for phantom
(self-intersecting) polymers $\xi=1/2$, and for self-avoiding polymers
$\xi=\nu$, with $\nu=3/4$ in two and $\approx0.588$ in three
dimensions respectively. Similarly, for polymer dynamics in the
absence of hydrodynamic interactions (Rouse: $\kappa=1+2\xi$
\cite{rouse,kay}), and polymer dynamics in a good solvent (Zimm
\cite{zimm,degennes,de}: $\kappa=3\xi$). The above means, in the
simplest case, that the mean-square displacement (MSD) of a tagged
monomer of a polymer must behave $\sim t^{2\xi/\kappa}$ until time
$\tau$, and $\sim t$ thereafter. Since $2\xi/\kappa$ is not
necessarily unity, the dynamics of a tagged monomer in a polymer is
anomalous till time $\tau$, and the polymer's diffusion coefficient
scales $\sim N^{2\xi-\kappa}$.

Starting from a microscopic description, there are two main approaches
to model anomalous dynamics in stochastic systems \cite{rainer}: (i)
Continuous Time Random Walk (CTRW) \cite{mont} and the associated
fractional Fokker-Planck Equation (fFPE), providing a probabilistic
description of phase-space trajectories, and (ii) the Generalized
Langevin Equation (GLE) \cite{morikubozwan}, which describes
individual phase space trajectories. Physical systems exhibiting
anomalous dynamics, for which probabilistic description of phase-space
trajectories as well as description of individual trajectories can be
obtained, are not only relatively rare, but also, relating one
description to the other often requires approximations
\cite{wang}. Anyhow, given the ubiquity of anomalous dynamics in
polymeric systems \cite{degennes,de}, one would expect them to have
been thoroughly examined from this perspective. To the best of my
knowledge however, probabilistic description of phase space
trajectories for anomalous polymer dynamics has only been considered
for isolated cases such as Rouse chain \cite{rouse1}, and polymer
translocation \cite{metz,vilgis,kantor,satya,vocksa}: in
\cite{metz,vilgis}, an fFPE approach has been put forward, wherein, a
(power-law) waiting time before each move of the monomer in the pore
is assumed to cause the anomalous dynamics. This approach is at odds
with numerical studies by others \cite{kantor}, who report that, for
the translocation of an infinite polymer, the probability distribution
is Gaussian in space, but with a width that scales anomalously in
time. Further, the anomalous dynamics of translocation has been shown
to match that of the fractional Brownian motion (fBm)
\cite{satya,vocksa}, which is also in contradiction with the fFPE
approach.

As for the description of individual trajectories for anomalous
polymer dynamics, two recent papers \cite{panja1,panja2} show that
without external forces, the motion of a tagged monomer of a polymer
--- for phantom and self-avoiding Rouse, Zimm, reptation,
translocation, and polymer melts --- is robustly described by a
unified Generalized Langevin Equation (GLE). The force $\vec\phi(t)$
experienced by the tagged monomer is related to its velocity $\vec
v(t)$ via
\begin{table*}
\begin{tabular}{c|c|c|c|c}
\hline\hline polymeric system & derivation of Eq. (\ref{e4})
\cite{de}& $\,\,\,\kappa\,\,\,$ & $\gamma_p$ & $k_p$\cr \hline\hline
phantom Rouse & directly from Rouse equation & $2$
&$\gamma_0=N\gamma$&$6\pi^2k_BTp^2/N$\cr &no approximation &
&$\gamma_{p\neq0}=2N\gamma$&\cr \hline Zimm (phantom) and &
$\,\,\,$Smoluchowski equation; pre-averaging$\,\,\,$ & $3/2$ &
$\,\,\,\gamma_0=3(6\pi^3N)^{1/2}\eta_s/8\,\,\,$ &$6\pi^2k_BTp^2/N$\cr
polymer in a $\theta$-solvent & approximation on the mobility matrix
&&$\gamma_{p\neq0}=(12\pi^3Np)^{1/2}\eta_s$&\cr\hline Zimm
(self-avoiding) & Smoluchowski equation; pre-averaging & $3\nu$
&$\gamma_0=\eta_s N^\nu$&$\,\,\, N^{-2\nu}k_BTp^{1+2\nu}\,\,\,$\cr &
approximation on the mobility matrix &&$\gamma_p=\eta_s N^\nu
p^{1-\nu}$&\cr \hline reptation & entropic (curvilinear) chain
tension; & $2$ &$\gamma_0=N\gamma$&$6\pi^2k_BTp^2/N$\cr
$\,\,\,$(curvilinear co-ordinate)$\,\,\,$ & i.e., Eq. (\ref{e4}) is
only one-dimensional &&$\gamma_{p\neq0}=2N\gamma$&\cr\hline\hline
\end{tabular}
\caption{Elaboration of Eq. (\ref{e4}) for phantom Rouse, Zimm,
  polymer in a $\theta$-solvent and reptation. Here $\eta_s$ is the
  solvent viscosity. Where applicable, the effective damping
  coefficient $\gamma$ for monomeric motion, as defined in
  Eq. (\ref{e2}), is $\propto\eta_s$. Note, in all cases, that
  $\kappa$ is simply the power of $p$ in the expression 
  $k_p/\gamma_p$ for $p\neq0$.\label{table1}}
\end{table*}
\begin{eqnarray}   
  \vec\phi(t)=-\int_0^tdt'\mu(t-t')\vec v(t')+\vec g(t),
\label{e1}
\end{eqnarray}   
where $\mu(t)$ is the memory kernel, and the noise $\vec g(t)$
satisfies $\langle\vec g(t)\rangle_0=0$ and the
fluctuation-dissipation theorem (FDT) $\langle
g_\sigma(t)g_\lambda(t')\rangle_0=k_BT\delta_{\sigma\lambda}\mu(t-t')$,
with $\sigma,\lambda=(x,y,z)$. Here $\langle\ldots\rangle_0$ denotes
an average over the noise realizations, including an average over
equilibrium configurations of the polymers at $t=0^-$. Further, with
$\vec v(t)$ responding to $\vec\phi(t)$ as $\gamma\vec
v(t)=\vec\phi(t)+\vec f(t)$, where $\gamma$ is the (effective) damping
coefficient for monomeric motion, and $\vec f(t)$ is a random force
satisfying $\langle\vec f(t)\rangle=0$ and the FDT $\langle
f_\sigma(t)f_\lambda(t')\rangle=2\gamma
k_BT\delta_{\sigma\lambda}\delta(t-t')$, one has
\begin{eqnarray}  
  \vec v(t)=\gamma^{-1}\left[-\int_0^tdt'\mu(t-t')\vec v(t')+\vec g(t)+\vec
  f(t)\right].
\label{e2}
\end{eqnarray} 
In this formulation, $\mu(t)\sim t^{-\alpha}e^{-t/\tau}$ for some
$0<\alpha<1$ (for a list, see Table I of \cite{panja2}). The FDT then
ensures that the MSD $\sim t^\alpha$ till time $\tau$, and
$\sim t$ thereafter. This formulation also robustly yields the correct
drift behavior of a tagged monomer under weak external forces
\cite{panja2}, like the Nernst-Einstein relation. The GLE
(\ref{e1}-\ref{e2}), describing non-Markovian trajectories in phase
space, demonstrate that there is no power-law waiting time (assumed in
the modeling of translocation by the fFPE); instead, the anomalous
dynamics stems from the fact that each move of the tagged monomer
tends to be undone later.

Here I demonstrate that, with a $\delta$-function distribution in 3D
space at $t=0$, the probability distribution of a tagged monomer for
phantom Rouse, self-avoiding Rouse, Zimm, polymers in a
$\theta$-solvent, reptation, translocation, and polymer melts in the
entangled regime is given by
\begin{eqnarray}
P(\vec r,t|\vec r_0,0)=e^{-(\vec r-\vec
r_0)^2/2\Delta(t)}/[2\pi\Delta(t)]^{3/2},
\label{e3}
\end{eqnarray}
with $\Delta(t)=At^\alpha e^{-t/\tau}+BN^{2\xi-\kappa}t$, where $A$
and $B$ are two system parameters-dependent constants, and
$N^{2\xi-\kappa}$ is the scaling of the polymer's diffusion
coefficient with $N$ [for translocation, Eq. (\ref{e3}) holds only
until the polymer disengages from the pore; see later]; i.e., the
anomalous dynamics of an infinite polymer is that of the fBm
\cite{mandel}. One should keep in mind that Eq. (\ref{e3}) is
demonstrated here for polymeric systems wherein the polymers are far
away from any boundary (including translocation), as Eq. (\ref{e3})
cannot be reconciled with nontrivial boundary conditions. The GLE
(\ref{e1}-\ref{e2}) and Eq. (\ref{e3}) thus provide a {\it complete\/}
description (i.e., of individual trajectories as well as that of
trajectory distribution in phase space) of anomalous dynamics in
polymeric systems; as mentioned earlier, this is relatively rare for
physical systems.

In fact, $P(\vec r,t|\vec r_0,0)$ for phantom Rouse, Zimm, polymers in
a $\theta$-solvent, and reptation can be obtained analytically, {\it
  irrespective\/} of the GLE description, thanks to the fact that
their dynamics is described by that of the polymer's fluctuation
modes. For these systems, the location $\vec R_n(t)$ of monomer $n$
can be expressed, in terms of the mode amplitude $\{\vec X_p(t)\}$s
for $p=0,1,2\ldots$, as $\vec R_n(t)=\vec X_0(t)+\sum_{p=1}^\infty\vec
X_p(t)\cos(\pi pn/N)$, obeying the boundary condition that the
polymer's chain tension vanishes at the free ends, i.e.,
$[\partial\vec R_n(t)/\partial n]_{n=0}=[\partial\vec R_n(t)/\partial
n]_{n=N}=0$. As the $\{\vec X_p(t)\}$s can be obtained by an inverse
cosine transformation of the above, the polymer dynamics is simply
reconstructed from the LE satisfied by each spatial component (denoted
by $\sigma$) of the $\{\vec X_p(t)\}$s \cite{de}
\begin{eqnarray}    
\gamma_p\dot X_{p\sigma}(t)=-k_p X_{p\sigma}(t)+f_{p\sigma}(t),
\label{e4}
\end{eqnarray}    
 where the stochastic force satisfies $\langle
f_{p\sigma}(t)\rangle=0$ and the FDT $\langle
f_{p\sigma}f_{q\lambda}\rangle=2\gamma_pk_BT\delta_{\sigma\lambda}\delta_{pq}\delta(t-t')$.
The list of $\gamma_p$ and $\kappa_p$ values for these systems appear
in Table \ref{table1}: note that the relaxation time $\tau_p$ for the
$p$-th mode $\tau_p=\gamma_p/k_p=c^{-1}p^{-\kappa}$ for $p\neq0$; one
can calculate $c$ from the table.
\begin{table}[h]
\begin{tabular}{c|c} \hline\hline polymeric
system&$\Delta(t)$\cr\hline\hline phantom Rouse & $\sim t^{1/2}$ till
$\tau$ and $\sim t$ thereafter\cr\hline Zimm (phantom) and & $\sim
t^{2/3}$ till $\tau$ and $\sim t$ thereafter\cr polymer in a
$\theta$-solvent&\cr\hline Zimm (self-avoiding) & $\sim t^{2/3}$ till
$\tau$ and $\sim t$ thereafter\cr\hline reptation & $\sim t^{1/2}$
till $\tau$ and $\sim t$ thereafter\cr (curvilinear
co-ordinate)&\cr\hline self-avoiding Rouse & $\sim
t^{\frac{2\nu}{1+2\nu}}$ till $\tau$ and $\sim t$
thereafter\cr\hline\hline
\end{tabular}
\caption{Systems of Table \ref{table1} and their
  $\Delta(t)$-behavior. Note that the scaling exponent of $\Delta(t)$ with
  time, in each case, is given by $2\xi/\kappa$, as noted in the
  first paragraph of the paper.\label{table2}}
\end{table}

\begin{figure}[h]
\begin{center}
\includegraphics[width=0.9\linewidth]{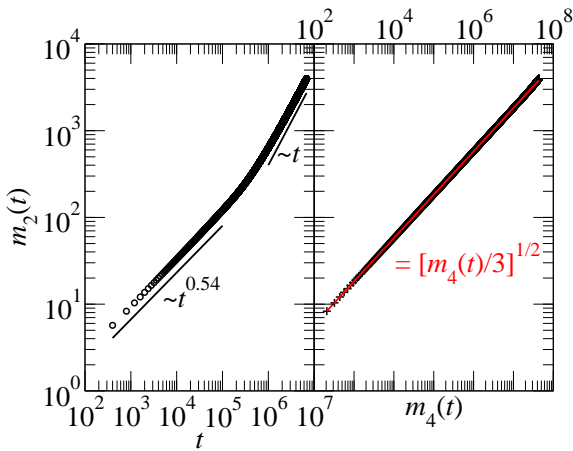}
\end{center}
\caption{(color online) Equation (\ref{e3}) for self-avoiding Rouse
  polymers ($N=400$): model details can be found in Appendix B. Left:
  second moment $m_2(t)
  =\sum_\sigma\langle[r_\sigma(t)-r_\sigma(0)]^2\rangle/3$ of the
  distribution (\ref{e3}), scales as $t^{\frac{2\nu}{1+2\nu}}\sim
  t^{0.54}$ till $\tau$, and $\sim t$ thereafter. Right: $m_2(t)$
  against
  $m_4(t)=\sum_\sigma\langle[r_\sigma(t)-r_\sigma(0)]^4\rangle/3$. Red
  line: $m_2(t)=[m_4(t)/3]^{1/2}$ --- note that $m_4(t)=3[m_2(t)]^2$
  for a Gaussian distribution. Data obtained from a time-series of
  200,000 consecutive snapshots, separated by 400 time units each, of
  256 different polymers.\label{fig1}}
\end{figure}

With the above, $P(\vec r,t|\vec r_0,0)$ is calculated as follows. For
fixed $\{\vec X_p^{(0)}\}$s that correspond to the position $\vec r_0$
of the tagged monomer at $t=0$, one determines the probability ${\cal
  P}(\vec X_p,t|\vec X_p^{(0)},0)$ from Eq. (\ref{e4}). One then
obtains $P(\vec r,t|\vec r_0,0)$ by integrating over all values of
$\{X_p(t)\}$ that correspond to the position $\vec r$ of the tagged
monomer at time $t$, taking into account the equilibrium distribution
of the $\{\vec X_p^{(0)}\}$-values. With the tagged monomer being the
middle monomer, this calculation, demonstrating Eq. (\ref{e3}) for
phantom Rouse, Zimm, polymers in a $\theta$-solvent, and reptation are
detailed in Appendix A: therein it is seen that
[c.f. Eqs. (\ref{s14}-\ref{s18})]
\begin{eqnarray}
\Delta(t)\sim\int_0^\infty \frac{dp}{k_p}\,\left(1-e^{-t/\tau_p}\right),
\label{add}
\end{eqnarray}
i.e., the exponent of $\Delta(t)$ in time is determined by the
integral in Eq. (\ref{add}). The corresponding scalings of $\Delta(t)$
summarized in Table \ref{table2}. Further, based on \cite{prouse}, I
note that one can construct an effective Eq. (\ref{e4}) also for
self-avoiding polymers, with $p$-independent $\gamma_0$ and
$\gamma_{p\neq0}$, and $k_p\sim p^{1+2\nu}$. The corresponding
scaling behavior for $\Delta(t)$, also calculated in Appendix A, is
listed in Table \ref{table2}, and is verified by simulations in
Fig. \ref{fig1}.

It is interesting to note here that the mode expansion technique to
establish Eq. (\ref{e3}) for the systems of Table \ref{table1} implies
that Eq. (\ref{e3}) is a simple consequence of the physical
connectivity of the polymer chain.

The expansion of the monomer co-ordinates into polymer's fluctuation
modes also provides an insight into how, given the GLE
(\ref{e1}-\ref{e2}), one may expect $P(\vec r,t|\vec r_0,0)$ to be
Gaussian. E.g., consider an ensemble ${\cal E}$ of polymers, at
equilibrium at $t=0^-$, with a given velocity history of the middle
monomer between times $0$ and $t$. For this ensemble, if $[\vec
g(t)+\vec f(t)]$ can be shown to be Gaussian, then the displacements
of the middle monomer in an infinitesimal time between $t$ and $(t+dt)$
is Gaussian distributed about the mean
$-\gamma^{-1}\int_0^tdt'\mu(t-t')\vec v(t')$, since $\vec v(t)$ is
proportional to $[\vec g(t)+\vec f(t)]$. Such (infinitesimal) Gaussian
displacements, accumulated over time, would then mean that
$P(r_\sigma,t|r_{\sigma0},0)$ has to be Gaussian.

In order to mathematically appreciate the Gaussian behavior of $[\vec
g(t)+\vec f(t)]$, one needs to recall the physics behind the anomalous
dynamics for these systems \cite{panja1,panja2}: a move of the middle
monomer creates a local strain by altering the polymer's chain tension
locally [Eq. (\ref{e1})]. In response to this strain, in subsequent
times there is an enhanced chance, for the monomer, to undo the move
[Eq. (\ref{e2})]. This physics is best represented by discretizing the
movement of the middle monomer in time for the ensemble ${\cal E}$
\cite{panja2}; e.g., $v(t)=\sum_{i=0}^m\delta\vec r_i\delta(t-t_i)$
for some $m$, and $t_0=0$ by choice. In between these moves the middle
monomer remains stationary, i.e., the dynamics of the polymer is given
by those of the mode amplitudes $\{\vec Y_p(t)\}$s, with the monomer
locations $\vec r_n(t)$ relative to the middle monomer expressed as
$\vec r_n(t)=4\sum_p\vec
Y^{(r)}_p(t)\sin\frac{\pi(2p+1)n}N\,\Theta(n-N/2)-4\sum_p\vec
Y^{(l)}_p(t)\sin\frac{\pi(2p+1)n}N\,\Theta(N/2-n)$ for
$p=0,1,2,\ldots$ (the superscripts for $\vec Y(t)$ correspond to the
right and the left halves of the polymer). The $\{Y_p(t)\}$s are
obtained from $\{\vec r_n(t)\}$s via the inverse sine transform, and
are readily shown to satisfy the boundary condition that the chain
tension vanishes at the open ends of the polymer. Through this
formulation, the polymer's chain tension at the middle monomer,
expressed in terms of the $\{\vec Y_p(t)\}$s, changes discretely at
$\{t_i\}$s, while in between, the relaxation of the chain tension
gives rise to the memory kernel of Eq. (\ref{e1}). To work this out
for all systems of Table \ref{table1}, one needs to re-perform, as
applicable, the pre-averaging approximation in terms of the $\{\vec
Y_p(t)\}$s, which is a cumbersome task. For the sake of simplicity, I
therefore only consider the phantom Rouse case here; for this system,
in times $t_i<t<t_{i+1}$, the $\{\vec Y_p(t)\}$s for each half {\it
independently\/} obey the LE \cite{panja2}
\begin{eqnarray}     \gamma_p\frac{\partial\vec Y_p}{\partial
t}=-q_p\,\vec Y_p(t)+\vec h_p(t),
\label{e5}
\end{eqnarray}      with $\gamma_p=2N\gamma$,
$q_p=6\pi^2k_BT(2p+1)^2/N$, $\tau_{2p+1}=\gamma_p/q_p$, $\langle\vec
h_p\rangle=0$ and $\langle
h_{p\sigma}(t)h_{q\lambda}(t')\rangle=\gamma_pk_BT\delta(t-t')\delta_{pq}\delta_{\sigma\lambda}$. Then
$\vec g(t)$ in Eq. (\ref{e1}-\ref{e2}) is given by (see Eq. (19) of
\cite{panja2})
\begin{eqnarray}     \vec
g(t)=4\sum_p\!\frac{\pi(2p+1)}N\Bigg\{\!e^{-t/\tau_{2p+1}}[\vec
Y^{(r)}_p(0^-)+\vec Y^{(l)}_p(0^-)]\nonumber\\
&&\hspace{-7.5cm}+\frac1\gamma_p\int_{0}^t \!\!dt'\,
e^{-(t-t')/\tau_{2p+1}}[\vec h^{(r)}_p(t')+\vec h^{(l)}_p(t')]\Bigg\}.
\label{e6}
\end{eqnarray}  With $\vec f(t)$, $\{\vec h_p(t)\}$s, $\{\vec
Y_p(0^-)\}$s being Gaussian distributed with zero mean [as of
Eqs. (\ref{e2}) and (\ref{e6}) respectively], $[\vec g(t)+\vec f(t)]$
also has to be Gaussian. 
\begin{figure*}
\begin{center}
\begin{minipage}{0.475\linewidth}
\includegraphics[width=\linewidth]{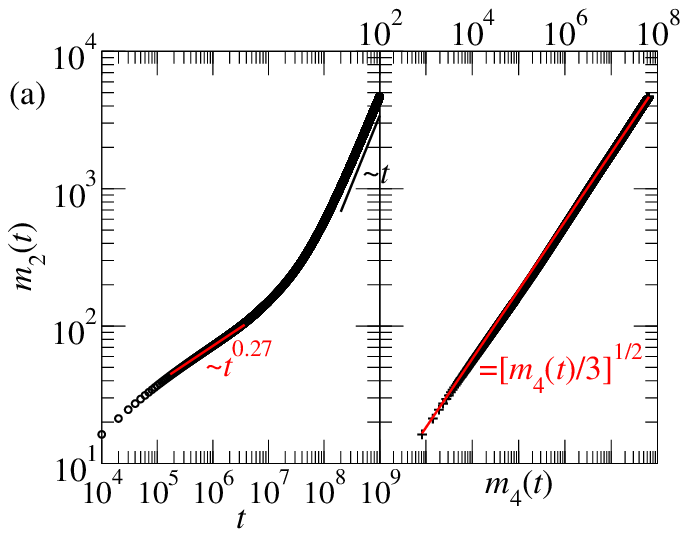}
\end{minipage}
\hspace{5mm}
\begin{minipage}{0.45\linewidth}
\includegraphics[width=\linewidth]{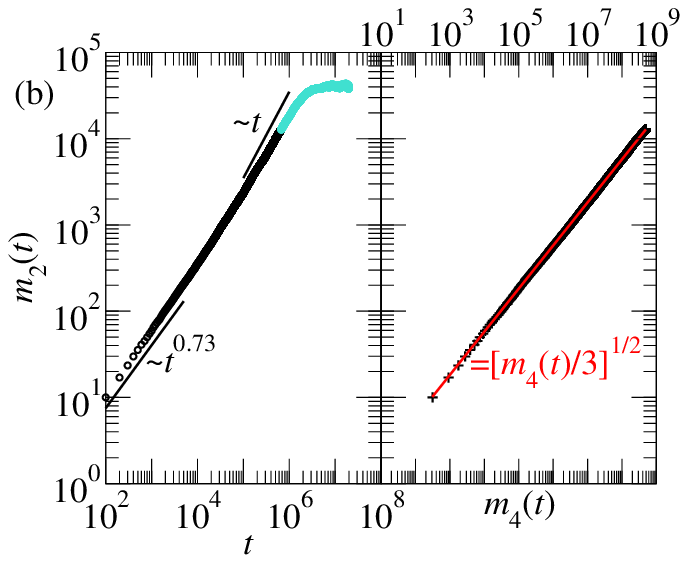}
\end{minipage}
\end{center}
\caption{(color online) (a) Left: The second moment
  $m_2(t)=(1/3)\sum_\sigma\langle[r_\sigma(t)-r_\sigma(0)]^2\rangle$
  of the distribution (\ref{e3}) for the middle monomer of a tagged
  polymer in a melt. Right: $m_2(t)$ against
  $m_4(t)=(1/3)\sum_\sigma\langle[r_\sigma(t)-r_\sigma(0)]^4\rangle$;
  red line: $m_2(t)=[m_4(t)/3]^{1/2}$. Data obtained from long
  time-series of 155,000 consecutive snapshots, separated by $10^4$
  time units each, of 1,728 different polymers with $N=1500$. See text
  for graph description. (b) Left: The second moment
  $m_2(t)=[s(t)-N/2]^2$, where $s(t)$ is the monomer number threaded
  in the pore at time $t$, $s(0)=N/2$. Right: $m_2(t)$ against
  $m_4(t)=\langle[s(t)-N/2]^4\rangle$; red line:
  $m_2(t)=[m_4(t)/3]^{1/2}$. Data averaged over 8,192 different
  polymers of $N=1000$, for the turquoise points in the left graph at
  least one polymer of the 8,192 polymers has disengaged from the
  pore. See text for graph descriptions. \label{fig2}}
\end{figure*}

Thus, to summarize so far, having expanded the monomer co-ordinates in
polymer's fluctuation modes, I have shown that $P(\vec r,t|\vec
r_0,0)$ for phantom Rouse, Zimm, polymers in a $\theta$-solvent,
reptation and self-avoiding Rouse is Gaussian; and have illustrated,
for the specific case of phantom Rouse, that Gaussianity of $P(\vec
r,t|\vec r_0,0)$ is expected from the GLE (\ref{e1}-\ref{e2}) since
the noise term $[\vec g(t)+\vec f(t)]$ is Gaussian. Unfortunately
however, the mode expansion does not work as simply for translocation
(for which, the tagged monomer is the one within the pore at time $t$,
i.e., it does not even have a fixed index) and for polymer
melt. Neither can Gaussianity of $[\vec g(t)+\vec f(t)]$ be shown, and
therefore, one has to rely on computer simulations. The data for the
melt are presented in Fig. \ref{fig2}(a): simulations are performed at
overall monomer density unity; the simulation details, same as that of
\cite{panja2} where the GLE has been shown to describe the dynamics of
the middle monomer of a tagged polymer in the entangled regime, can be
found in Appendix B. Following reptation theory for polymer melts, one
expects the second moment $m_2(t)$ to behave $\sim t^{1/4}$ in the
entangled regime, which starts around time $\sim 10^5$ for this model;
in this regime an effective exponent 0.27 is found (the red line in
the left graph). While the right graph is consistent with
Eq. (\ref{e3}) in the entangled regime, between the entangled regime
and the eventual diffusive regime, $P(\vec r,t|\vec r_0,0)$ does
deviate very slightly, and systematically, from (\ref{e3}), indicating
that the noise term $[\vec g(t)+\vec f(t)]$ is not Gaussian during
this time. 

The data for unbiased translocation are presented in
Fig. \ref{fig2}(b) [see model details in Appendix B]. The monomer
number within the pore at time $t$ is denoted by $s(t)$. Polymers are
equilibrated with $s(0)=N/2$. Here $m_2(t)\sim
t^{(1+\nu)(1+2\nu)}\approx t^{0.73}$, and should cross over to
diffusive behavior, as predicted in \cite{anom}, although the
crossover is slow: at long times polymers start disengaging from the
pore, corresponding to the flattening of $m_2(t)$ (the turquoise
points in the left graph), hence the true diffusive behavior can only
be observed for very long polymers. Nevertheless, Eq. (\ref{e3}) is
verified cleanly up to the point when all polymers remain threaded in
the pore: this behavior is in agreement with \cite{kantor,satya}
[albeit they report an anomalous exponent different from
$(1+\nu)/(1+2\nu)$], and contradicts \cite{vilgis}, reaffirming that
fFPE is not applicable for polymer translocation.

Finally, I note that the drift of a tagged monomer due to a weak
external force $\vec F$ (i.e., in the linear response regime), is
given by $t^\alpha \vec F$ till time $\tau$ and $t\vec F$ thereafter,
where $\alpha$ is the anomalous exponent. Describing this requires a
simple extension of Eq. (\ref{e3}). 

Ample computer time from the Dutch national supercomputer facility
SARA, and help from Gerard Barkema with simulations are gratefully
acknowledged. It is also a pleasure to thank Gerard Barkema and Rainer
Klages for helpful comments on the manuscript.

\appendix

\setcounter{equation}{0}
\renewcommand{\theequation}{A\arabic{equation}}

\begin{widetext}

\section{Appendix A: : Derivation of Eq. (\ref{e3}) for the middle monomer for
phantom Rouse, Zimm, polymers in a $\theta$-solvent, reptation, and
self-avoiding Rouse polymers}

I tag the middle monomer of the polymer, and here I obtain
Eq. (\ref{e3}) for its dynamics.

I start with the Langevin equation (\ref{e3}) describing the evolution
of the $p$-th mode amplitude ($p=0,1,2,\ldots$), viz.,
\begin{eqnarray}  \gamma_p\frac{X_{p\sigma}(t)}{\partial t}=-k_p
X_{p\sigma}(t)+f_{p\sigma}(t)\,,
\label{s1}
\end{eqnarray}  with $\langle f_{p\sigma}\rangle=0$ and the FDT
$\langle
f_{p\sigma}(t)f_{q\lambda}(t')\rangle=2\gamma_pk_BT\delta(t-t')\delta_{pq}\delta_{\sigma\lambda}$.
As noted in Table \ref{table1}, Eq. (\ref{s1}) can be derived for
phantom Rouse, Zimm, polymers in a $\theta$-solvent and reptation. A
straightforward result that follows from Eq. (\ref{s1}) is that
\begin{eqnarray}  \langle
X_{p\sigma}(t)X_{q\lambda}(t')\rangle=\delta_{pq}\delta_{\sigma\lambda}(k_BT/k_p)e^{-k_p(t-t')/\gamma_p}=\delta_{pq}\delta_{\sigma\lambda}(k_BT/k_p)e^{-(t-t')/\tau_p},
\label{s2}
\end{eqnarray}  where $\tau_p$ is the relaxation time of the $p$-th
mode ($p\neq0$) for the polymer. When Eq. (\ref{s2}) is combined with
the corresponding time correlation function for mode amplitudes for
self-avoiding polymers \cite{prouse}, namely $\langle\vec
X_p(t)\cdot\vec X_q(t')\rangle\propto
N^{2\nu}p^{-(1+2\nu)}e^{-t/\tau_p}\delta_{pq}$ with
$\tau_p\sim(N/p)^{1+2\nu}$, one can formulate an effective
Eq. (\ref{e4}) with both $\gamma_0$ and $\gamma_{p\neq0}$ independent
of $p$, and $k_p\sim p^{-(1+2\nu)}$ [this implies that for a self
avoiding polymer $\tau_p\sim(N/p)^{1+2\nu}$]. Given this, I will
henceforth use Eq. (\ref{s1}) also for self-avoiding Rouse polymers.

The Fokker-Planck equation for the probability ${\cal
P}(X_{p\sigma},t)$ that corresponds to the LE is given by
\cite{vankampen}
\begin{eqnarray}  \frac{\partial {\cal P}(X_{p\sigma},t)}{\partial
t}=\underbrace{\frac{k_p}{\gamma_p}}_{=\tau^{-1}_p\,\,\mbox{for}\,\,p\neq0}\frac{\partial}{\partial
X_{p\sigma}}\left[X_{p\sigma}{\cal
P}(X_{p\sigma},t)\right]+\underbrace{\frac{k_BT}{\gamma_p}}_{=a_p\tau^{-1}_p\,\,\mbox{for}\,\,p\neq0}\frac{\partial^2
{\cal P}(X_{p\sigma},t)}{\partial X^2_{p\sigma}},
\label{s3}
\end{eqnarray}  where $a_p=k_BT/k_p$ for $p\neq0$, and
$a_0=k_BT/\gamma_0$. The solution of Eq. (\ref{s3}), with the initial
condition that
$P(X_{p\sigma},0)=\delta(X_{p\sigma}-X^{(0)}_{p\sigma})$, is obtained
as follows.
\begin{itemize}
\item[(i)] For $p=0$, $k_p=0$ i.e., (\ref{s3}) is a simple diffusion
equation. Its solution is given by \cite{uh}
\begin{eqnarray} {\cal P}(X_{0\sigma},t)=\frac{1}{\sqrt{2\pi
a_0t}}\,\exp\left[-\frac{(X_{0\sigma}-X^{(0)}_{0\sigma})^2}{2a_0t}\right].
\label{s4}
\end{eqnarray}
\item[(ii)] For $p\neq0$, Eq. (\ref{s3}) can be verified by direct
substitution of its solution
\begin{eqnarray}  {\cal P}(X_{p\sigma},t)=\frac{1}{\sqrt{2\pi
a_p(1-e^{-2t/\tau_p})}}\,\exp\left[-\frac{(X_{p\sigma}-X^{(0)}_{p\sigma}e^{-t/\tau_p})^2}{2a_p(1-e^{-2t/\tau_p})}\right].
\label{s5}
\end{eqnarray}
\end{itemize} Next, as noted above Eq. (\ref{e4}), in terms of the
mode amplitudes, the location of the middle monomer ($n=N/2$) at any
time $t$ is given by
\begin{eqnarray}  \vec r(t)=\vec X_0(t)+2\sum_{p=1}^\infty\vec
X_p(t)\,\cos\frac{p\pi}{2}.
\label{s6}
\end{eqnarray}  Using Eq. (\ref{s6}), I obtain, upon averaging over
all possible initial states of the polymer at $t=0$
\begin{eqnarray}  P(r_\sigma,t|r_{0\sigma},0)=\prod_{p=0}^\infty
\int_{-\infty}^\infty dX^{(0)}_{p\sigma}\,{\cal
P}_{\text{eq}}(X^{(0)}_{p\sigma})\,\,\delta\!\!\left[r_{0\sigma}-\left(X^{(0)}_{0\sigma}+2\sum_{q=1}^\infty
X^{(0)}_{q\sigma}\,\cos\frac{q\pi}{2}\right)\right]\nonumber\\&&\hspace{-10cm}\times\int_{-\infty}^\infty
dX_{p\sigma}\,{\cal
P}(X_{p\sigma},t)\,\,\delta\!\!\left[r_\sigma-\left(X_{0\sigma}+2\sum_{q=1}^\infty
X_{q\sigma}\,\cos\frac{q\pi}{2}\right)\right],
\label{s7}
\end{eqnarray} where ${\cal P}_{\text{eq}}(X)$ is the equilibrium
probability of $X$, i.e., a Gaussian, obtained by taking the
$t\rightarrow\infty$ limit of Eq. (\ref{s5}).

At this stage, because of the $\delta$-functions in Eq. (\ref{s7}), it
is easiest to Fourier transform $P(r_\sigma,t|r_{0\sigma},0)$, defined
as
\begin{eqnarray}  \tilde{\cal P}_{k,k';t}=\frac1{2\pi}\int dr_\sigma\,
dr_{0\sigma} e^{i\left[k'r_\sigma+kr_{0\sigma}\right]}\,{\cal
P}(r_\sigma,t|r_{0\sigma},0],
\label{s8}
\end{eqnarray}  which reduces Eq. (\ref{s7}) to
\begin{eqnarray}
2\pi\,e^{-i\left[kr_\sigma+k'r_{0\sigma}\right]}\tilde {\cal
P}_{k,k';t}=\prod_{p=0}^\infty \int_{-\infty}^\infty
dX^{(0)}_{p\sigma}\,{\cal
P}_{\text{eq}}(X^{(0)}_{p\sigma})\,e^{-ik\left[X^{(0)}_{0\sigma}+2\sum_{q=1}^\infty
X^{(0)}_{q\sigma}\,\cos\frac{q\pi}{2}\right]}\nonumber\\&&\hspace{-8cm}\times\int_{-\infty}^\infty
dX_{p\sigma}\,{\cal
P}(X_{p\sigma},t)\,e^{-ik'\left[X_{0\sigma}+2\sum_{q=1}^\infty
X_{q\sigma}\,\cos\frac{q\pi}{2}\right]},
\label{s9}
\end{eqnarray}  At this point, in order to follow through the
calculation of $\tilde {\cal P}_{k,k';t}$, I need the two following
integrals:
\begin{itemize}
\item[(a)]
\begin{eqnarray}  \int_{-\infty}^\infty
dX_{0\sigma}\,\frac{1}{\sqrt{2\pi
a_0t}}\,\exp\left[-\frac{(X_{0\sigma}-X^{(0)}_{0\sigma})^2}{2a_0t}-ik'X_{0\sigma}\right]=e^{-iX^{(0)}_{0\sigma}k'-\frac12a_0tk'^2}.
\label{s10}
\end{eqnarray}
\item[(b)] for $p\neq0$:
\begin{eqnarray}  \int_{-\infty}^\infty
dX_{p\sigma}\,\frac{1}{\sqrt{2\pi
a_p(1-e^{-2t/\tau_p})}}\,\exp\left[-\frac{(X_{p\sigma}-X^{(0)}_{p\sigma}e^{-t/\tau_p})^2}{2a_p(1-e^{-2t/\tau_p})}-2ik'X_{p\sigma}\cos\frac{p\pi}2\right]\nonumber\\&&\hspace{-10cm}=e^{-2iX^{(0)}_{p\sigma}e^{-t/\tau_p}k'\cos(p\pi/2)-2a_p(1-e^{-2t/\tau_p})k'^2\cos^2(p\pi/2)}.
\label{s11}
\end{eqnarray}
\end{itemize} Using (a-b), I now integrate over $X^{(0)}_{0\sigma}$
(i.e., the location the center-of-mass of the polymer) with a uniform
probability density measure yields $\sqrt{2\pi}\delta(k+k')$, which
leads me to
\begin{eqnarray}
\hspace{-5mm}\sqrt{2\pi}\,e^{-i\left[k'r_\sigma+kr_{0\sigma}\right]}\tilde
{\cal P}_{k,k';t}=e^{-k^2\left[\frac12a_0t+2\sum_{q=1}^\infty
a_q(1-e^{-2t/\tau_q})\cos^2(q\pi/2)\right]}\delta(k+k')\nonumber\\&&\hspace{-7.5cm}\times
\prod_{p=1}^\infty\!\!\left[ \int_{-\infty}^\infty\!\!\!\!\!
dX^{(0)}_{p\sigma}\,{\cal
P}_{\text{eq}}(X^{(0)}_{p\sigma})e^{-2ik\left[X^{(0)}_{p\sigma}(1-e^{-t/\tau_p})\cos(p\pi/2)\right]}\right]\nonumber\\&&\hspace{-8.1cm}=e^{-k^2\left[\frac12a_0t+2\sum_{q=1}^\infty
a_q\cos^2(q\pi/2)\{(1-e^{-2t/\tau_q})+(1-e^{-t/\tau_q})^2\}\right]}\delta(k+k')
\nonumber\\&&\hspace{-8.1cm}=e^{-k^2\left[\frac12a_0t+4\sum_{q=1}^\infty
a_q\cos^2(q\pi/2)(1-e^{-t/\tau_q})\right]}\delta(k+k');
\label{s13}
\end{eqnarray}  ${\cal P}_{k,k';t}\propto\delta(k+k')$ implies that
$P(r_\sigma,t|r_{0\sigma},0)$ is a function of
$(r_\sigma-r_{0\sigma})$.

Finally, I now need to evaluate the discrete sum in the exponent of
Eq. (\ref{s13}). Having noticed that $\cos(q\pi/2)=0$ for odd
$q$-values and $\cos^2(q\pi/2)=1$ for even $q$-values, the sum can be
converted into an integral; thereafter the inverse Fourier transform
from $k$ to $(r_\sigma-r_{0\sigma})$ leads to Eq. (\ref{e3}), with the
behavior of $\Delta(t)$ presented in Table \ref{table2}. With the
corresponding scaling of $\gamma_p$ and $\tau_p=\gamma_p/k_p$ for
phantom Rouse, Zimm, polymers in a $\theta$-solvent, reptation, and
self-avoiding Rouse polymers (see Table \ref{table1}), these integrals
are listed below. Note that in Eqs. (\ref{s14}-\ref{s18}) I omit
constants in converting the discrete sums to integrals.
\begin{itemize}
\item[A.] Phantom Rouse:
\begin{eqnarray} 4\sum_{q=1}^\infty
a_q\cos^2(q\pi/2)(1-e^{-t/\tau_q})\rightarrow k_BT\int_0^\infty
\frac{dq}{q^2}\,(1-e^{-cq^2t})\sim\sqrt t.
\label{s14}
\end{eqnarray}
\item[B.] Phantom Zimm and polymers in a $\theta$-solvent:
\begin{eqnarray} 4\sum_{q=1}^\infty
a_q\cos^2(q\pi/2)(1-e^{-t/\tau_q})\rightarrow k_BT\int_0^\infty
\frac{dq}{q^2}\,(1-e^{-cq^{3/2}t})\sim t^{2/3}.
\label{s15}
\end{eqnarray}
\item[C.] (self-avoiding) Zimm:
\begin{eqnarray} 4\sum_{q=1}^\infty
a_q\cos^2(q\pi/2)(1-e^{-t/\tau_q})\rightarrow k_BT\int_0^\infty
\frac{dq}{q^{1+2\nu}}\,(1-e^{-cq^{3\nu}t})\sim t^{2/3}.
\label{s16}
\end{eqnarray}
\item[D.] reptation (curvilinear co-ordinate):
\begin{eqnarray} 4\sum_{q=1}^\infty
a_q\cos^2(q\pi/2)(1-e^{-t/\tau_q})\rightarrow k_BT\int_0^\infty
\frac{dq}{q^2}\,(1-e^{-cq^2t})\sim\sqrt t.
\label{s17}
\end{eqnarray}
\item[E.] self-avoiding Rouse:
\begin{eqnarray} 4\sum_{q=1}^\infty
a_q\cos^2(q\pi/2)(1-e^{-t/\tau_q})\rightarrow k_BT\int_0^\infty
\frac{dq}{q^{1+2\nu}}\,(1-e^{-cq^{1+2\nu}t})\sim t^{2\nu/(1+2\nu)}.
\label{s18}
\end{eqnarray}
\end{itemize}  Clearly, these power-law behavior of $\Delta(t)$ cannot
hold longer than time $\tau$, this is also noted in Table
\ref{table2}.

\end{widetext}

\section{Appendix B:  Simulation details}

Over the past years, a highly efficient simulation approach to polymer
dynamics has been developed in our group. This is made possible via a
lattice polymer model, based on Rubinstein's repton model \cite{rub}
for a single reptating polymer, with the addition of sideways moves
(Rouse dynamics). A detailed description of this model, its
computationally efficient implementation and a study of some of its
properties and applications can be found in \cite{heuk1}.

In this model, each polymer is represented by a sequential string of
monomers, living on a face-centered-cubic lattice with periodic
boundary conditions in all three spatial directions. Hydrodynamic
interactions between the monomers are not taken into account in this
model. Monomers adjacent in the string are located either in the same,
or in neighboring lattice sites. The polymers are self-avoiding:
multiple occupation of lattice sites is not allowed, except for a set
of adjacent monomers. The number of stored lengths within any given
lattice site is one less than the number of monomers occupying that
site. The polymers move through a sequence of random single-monomer
hops to neighboring lattice sites. These hops can be along the contour
of the polymer, thus explicitly providing reptation dynamics. They can
also change the contour ``sideways'', providing Rouse dynamics. Each
kind of movement is attempted with a statistical rate of unity, which
defines the unit of time. This model has been used before to simulate
the diffusion and exchange of polymers in an equilibrated layer of
adsorbed polymers \cite{wolt1}, dynamics self-avoiding Rouse polymers
\cite{prouse1}, polymer translocation under a variety of circumstances
\cite{vocksa,anom,panjatrans}, and the dynamics of polymer adsorption
\cite{adsorb}.

The same model has been used for the polymer melt simulations (here
the polymers are both self- and mutually-avoiding) for a system of
size $60^3$ with an overall monomer density unity per lattice
site. Due to the possibility that adjacent monomers belonging to the
same polymer can occupy the same site, overall approximately 40\% of
the sites typically remain empty.

Initial thermalizations were performed as follows: completely crumpled
up polymers are placed in lattice sites at random. The system is then
brought to equilibrium by letting it evolve up to $10^9$ units of
time, with a combination of random intermediate redistribution of
stored lengths within each polymer. Additional details on the melt
simulations can be found in \cite{panja2}.

\end{document}